\documentstyle[prl,aps]{revtex}
\def\dfrac#1#2{{\displaystyle{#1\over#2}}}

\begin{document}
\twocolumn[\hsize\textwidth\columnwidth\hsize\csname
  @twocolumnfalse\endcsname
\title{A Fermi Fluid Description of the Half-Filled Landau Level}
\author{Tapash Chakraborty$^\star$}
%\author{Tapash Chakraborty}
\address{Max-Planck Institut f\"ur Physik Komplexer Systeme, 
N\"othnitzer Stra{\ss}e 38, D-01187 Dresden, Germany}
\date{\today}
\maketitle
\begin{abstract}
We present a many-body approach to calculate the ground state properties
of a system of electrons in a half-filled Landau level. Our starting point
is a simplified version of the recently proposed trial wave function where
one includes the antisymmetrization operator to the bosonic Laughlin state. 
Using the classical plasma analogy, we calculate the pair-correlation function, 
the static structure function and the ground state energy in the thermodynamic 
limit. These results are in good agreement with the expected behavior at 
$\nu=\frac12$.
\end{abstract}
  \vskip 2pc ] % end \twocolumn[...]

%\pacs{73.40.Hm,73.20.Dx,72.15.Nj}
\narrowtext

The fractional quantum Hall effect (FQHE) \cite{fqhe} which is understood
to be due to condensation of electrons to unique incompressible states as a 
result of electron correlations \cite{laughlin,book}, fails to explain the odd
behavior of even-denominator filling fractions which lie right in the middle of 
all the observed FQHE filling factors. It has been experimentally established 
that at $\nu=\frac12$ the system is metallic. (Here $\nu=\phi_0n_e/B$, where 
$\phi_0=hc/e$, $n_e$ is the mean electron density, and $B$ is externally 
applied magnetic field.) The seemingly metallic behavior observed in transport 
measurements \cite{jiang} was confirmed in subsequent surface acoustic wave 
experiments where it was found that contrary to the case of odd-denominator 
filling factors where the conductivity is reduced, at the half-filled Landau 
level the conductivity is in fact, enhanced \cite{willett}. Earlier theoretical 
attempts to understand the nature of the $\nu=\frac12$ state, largely pioneered 
by Halperin \cite{bert} and later by others \cite{haldane,fano,tapash} remained 
mostly inconclusive. While it was known from those theoretical works that the 
$\frac12$-state is compressible, the exact nature of the state remained unclear.
For example, working with upto ten electrons in a periodic rectangular geometry 
and the exact diagonalization of the few-electron Hamiltonian in the lowest 
Landau level, Haldane \cite{haldane} found that the excitation spectrum is 
particle number dependent, the ground state energy was never at the zero total 
momentum (contrary to what one expects in a uniform-density liquid), and no 
clear physical picture could be extracted from those numerical results. The 
ground state energy (in fact, the lowest energy) was also dependent on the 
electron number and extrapolation of the energies to an infinite system led to 
$E_0\approx-0.465 e^2/\epsilon\ell_0$ \cite{book} (here $\ell_0=\left(\hbar 
c/eB\right)^{\frac12}$ is the magnetic length). The Laughlin wave function 
\cite{laughlin}
\begin{equation}
\psi_L=\prod_{i<j}\left(z_i-z_j\right)^m\,\exp\left\{-\sum_k|z_k|^2/4
\ell_0^2\right\}
\end{equation}
where $z=x+iy$ is the electron position and $\nu=1/m$, describes a system of 
particles obeying Bose statistics for $m=2$ and can not be used for the
fermion system without any further modification.

In order to explain the anomalous results at $\nu=\frac12$, a very intriguing 
theory was proposed by Halperin, Lee, and Read (HLR) \cite{halperin}.
This theory describes the compressible even-denominator states in terms of
a transformation that represents each electron as a Chern-Simons fermion
carrying even number of fictitious magnetic flux quanta pointing in the
direction opposite to the external magnetic field. In a mean field
approximation (no interparticle interaction) the average fictitious field
cancels the real magnetic field and as a result the transformed fermions
experience no net field. They then form a gapless Fermi liquid. Subsequent
experimental observation \cite{fermisurf} of the geometric resonance of the 
quasiparticle cyclotron orbits with the acoustic waves, and similar geometrical 
resonances found in antidot arrays, indicated the existence of a Fermi surface 
at $\nu=\frac12$. These experiments provided strong support for the theoretical
picture of HLR. However, fluctuations beyond the mean field theory, which is 
expected just to renormalize the Fermi liquid parameters, instead found to 
cause divergences and the situation has not improved much from there yet 
\cite{stern}.

Parallel to the above approach, there is an ongoing effort to develop a
microscopic approach to $\nu=\frac12$ based on the idea of having an
improved Laughlin-like wave function as a starting point. One way to do
that is to include the antisymmetrization operator to the Laughlin state
and have a trial wave function \cite{duncan,rezayi}
\begin{equation}
\Psi={\cal P}_{LLL}\,\det M \prod_{i<j}\left(z_i-z_j\right)^2\,
\exp\left\{-\sum_k|z_k|^2/4\ell_0^2\right\}.
\end{equation}
Here ${\cal P}_{LLL}$ is the lowest Landau level projection operator.
The matrix $M$ has elements that are plane waves, $M_{ij}=e^{i{\bf k}_i
\cdot{\bf r}_j}$, $|{\bf k}|<k_F$. For $\nu=\frac12$, the Fermi wave vector
is $k_F = \left[4\pi n_e/s\right]^{\frac12}= 1/(\sqrt s\ell_0)$ where $s$ is 
the spin degeneracy. For a fully spin-polarized system $s=1$. Because of the 
projection operator, $\bar{z}_i\rightarrow2\dfrac{\partial}{\partial
z_i}$, and therefore the plane wave factors act as operators on the Jastrow
factor where, as a result, the zeroes of $\psi_L$ are displaced 
\cite{duncan,rezayi}. 
The wave function (2) is supposed to have the right statistics and right
correlations to describe the Fermi liquid properties at $\nu=\frac12$, and
is found to provide a good description of a small size system at $\nu=\frac12$ 
\cite{rezayi}. However, in those numeriucal studies of the few-electron systems
the ``Fermi'' pair-correlation function on a sphere was found to have distinct 
long-range type oscillations unlike the dominant short-range order present in 
a fluid and also not present in the Laughlin (``Boson'') state. Further, it is 
reasonable to question the reliability of a few-electron system result when we 
are to describe a gapless Fermi liquid. The pair-correlation function and the 
structure function for the state (2) are the most essential building blocks for 
any further development in the theory of a compressible fluid. The nature of 
the correlation functions in the thermodynamic limit, the effective mass, and 
collective excitations, which are related to those correlation functions, 
therefore need careful attention \cite{duncan}.

In this work, we have attempted to fill in for some of those open questions 
by appealing to the ingenuity of the original Laughlin approach where
one is able to map the electrons onto a classical plasma and make use of the
established formalism to calculate various physical quantities. To develop
such a many-body scheme to deal with the wave function (2) which includes
the projection operator is however a non-trivial problem. In our approach,
we start with a simplified problem and drop the projection operator from (2): 
the wave function is then a simple product of the Slater determinant and the 
Laughlin function for the $\frac12$-state
\begin{equation}
\Psi_F=\det M\psi_L=\Phi\psi_L.
\end{equation}
Our justification for that somewhat radical step is that, in our choice of the 
trial wave function $\Psi_F$ the only job of the Slater determinant $\Phi$ is 
to make $\Psi_F$ antisymmetric. When $\Phi=1$, the wave function describes the 
boson (Laughlin) fluid and when $\psi_L=1$ the wave function describes the 
non-interacting Fermi system. Although we have no longer any explicit
projection to the lowest Landau level, the Laughlin wave function, in
particular the analytic part of the wave function already describes the 
correlations in the lowest Landau level. Also, since we are interested 
primarily in the correlation functions, structure functions etc., it should 
perhaps still be an acceptable step to drop the projection operator, especially 
since the form of $|\Phi|^2$ is chosen to be of the same form as $|\psi_L|^2$ 
(see below). We wish to add here that for a Fermi liquid in the {\it absence} 
of a magnetic field, a division of labor as for the two functions in (3) is 
entirely justified. 

Once the choice of the wave function (3) is made the next question is, how
do we deal with $\Phi$. We have already stated that we are mostly interested 
in the pair-correlation functions where information about $|\Phi|^2$ is all
that need to be known, or more specifically, we need to constrain $|\Phi|^2$ 
to be positive definite. One available choice in the literature \cite{springer} 
which is quite successful in describing the correlated electron systems in
the absence of an external magnetic field is to write
\begin{eqnarray}
\sum_\sigma|\Phi({\bf r}_1,\ldots,{\bf r}_N)|^2 &\approx& \prod_{i<j}
\phi^2(r_{ij})\nonumber \\
\phi(r) &=& e^{-\frac12 u_I(r)}
\end{eqnarray}
where the set of spin coordinates is denoted by $\sigma$. This means that we
expand $|\Phi|^2$ and retain only the two-body term which is then approximated 
by a Jastrow-type function. This allows us to write the square of the total 
wave function as
\begin{eqnarray}
\left|\Psi\left({\bf r}_1,\cdots,{\bf r}_N\right)\right|^2 &=& \prod_{i<j}
e^{-\left[u(r_{ij})+u_I(r_{ij})\right]}\nonumber \\
&=& \prod_{i<j}\,e^{-u_t(r_{ij})}.
\end{eqnarray}
and the corresponding pair-correlation function 
\begin{eqnarray}
&g(r_{12})&=N(N-1)n_e^2\int d^2r_3\cdots d^2r_N\nonumber \\
&\times&\exp\left(-\sum_{i<j}u_t(r_{ij})\right)/\int d^{2N}r\exp\left(-
\sum_{i<j}u_t(r_{ij})\right)\nonumber \\
\end{eqnarray}
where $N$ is the particle number.

\begin{center}
\begin{picture}(200,200)
\put(0,0){\includegraphics{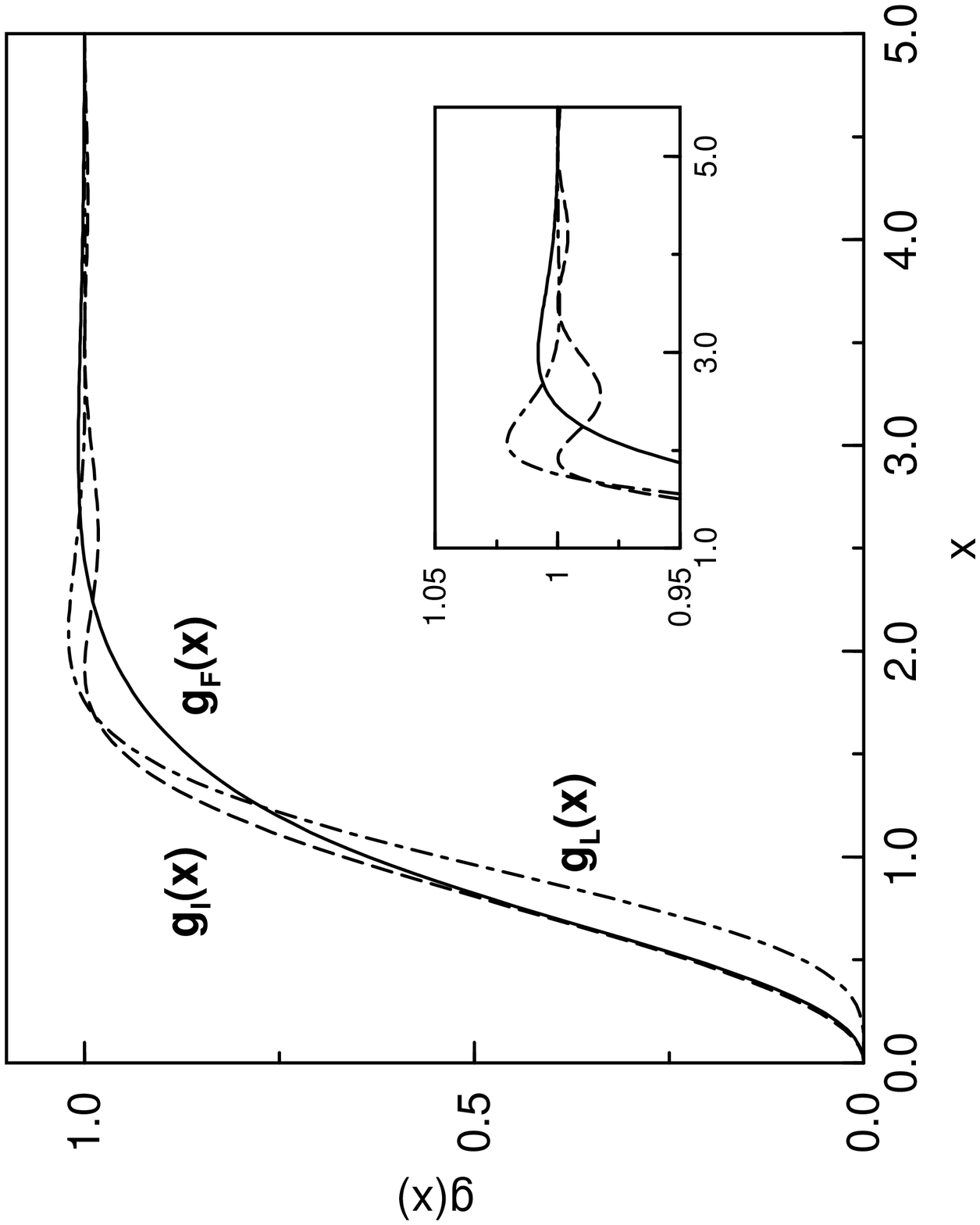}}
\label {fig1}
\end{picture}
\end{center}

\noindent
FIG. 1: The pair-correlation function $g(x)$ as a function of $x=r/R,
R=\sqrt{2m}\ell_0$ for the non-interacting system $[g_I(x)]$ [Eq. (7)], the 
Laughlin state (1) for a Bose system $(g_L(x)]$, and the ``Fermi'' state (3) 
$g_F(x)$. A blowup of the region around unity is given as inset.
\vskip 1.0 true cm

The advantage of our choice of (5) is that one can now use established methods 
like the celebrated mapping of Laughlin's wave function (1) to a one-component
classical plasma \cite{book,plasma} which determines the $u(r)$ or equivalently 
the pair-correlation function $g(r)$. In order to perform similar calculations 
for $u_t(r)=u(r)+u_I(r)$, we first have to determine $u_I(r)$ and then follow 
the plasma analogy to solve for $u_t(r)$. 
In a completely degenerate ideal (non-interacting) two-dimensional Fermi system 
the exact two-body radial distribution function $g_I(r)$ can be calculated to 
be \cite{campbell}
\begin{equation}
g_I(r) = 1-\left[2J_1(k_Fr)/k_Fr\right]^2,
\end{equation}
where $J_1(k_Fr)$ is the Bessel function of the first kind of order one. Since 
we are considering a fully spin polarized system, $g_I(r)$ vanishes at the 
origin due to the Pauli exclusion principle (Fig. 1). The corresponding ideal 
gas static structure function obtained from the two-dimensional Fourier
transform of (7) is \cite{campbell}
\begin{equation}
S_I(\kappa)= \left\{ \begin{array}{ll}
   \dfrac2\pi\left[\sin^{-1}\kappa+\kappa\left(1-\kappa^2\right)^{\frac12}
   \right],
   & \mbox{$\kappa < 1$} \nonumber \\
   1, & \mbox{$\kappa > 1$} \nonumber \\
\end{array}
\right.
\end{equation}
where $\kappa=k/2k_F$. For small $\kappa$, $S_I(\kappa)$ increases linearly 
with $\kappa$ (Fig. 2).  

\vspace*{-1.0cm}
\begin{center}
  \begin{picture}(200,200)
  \put(0,0){\includegraphics{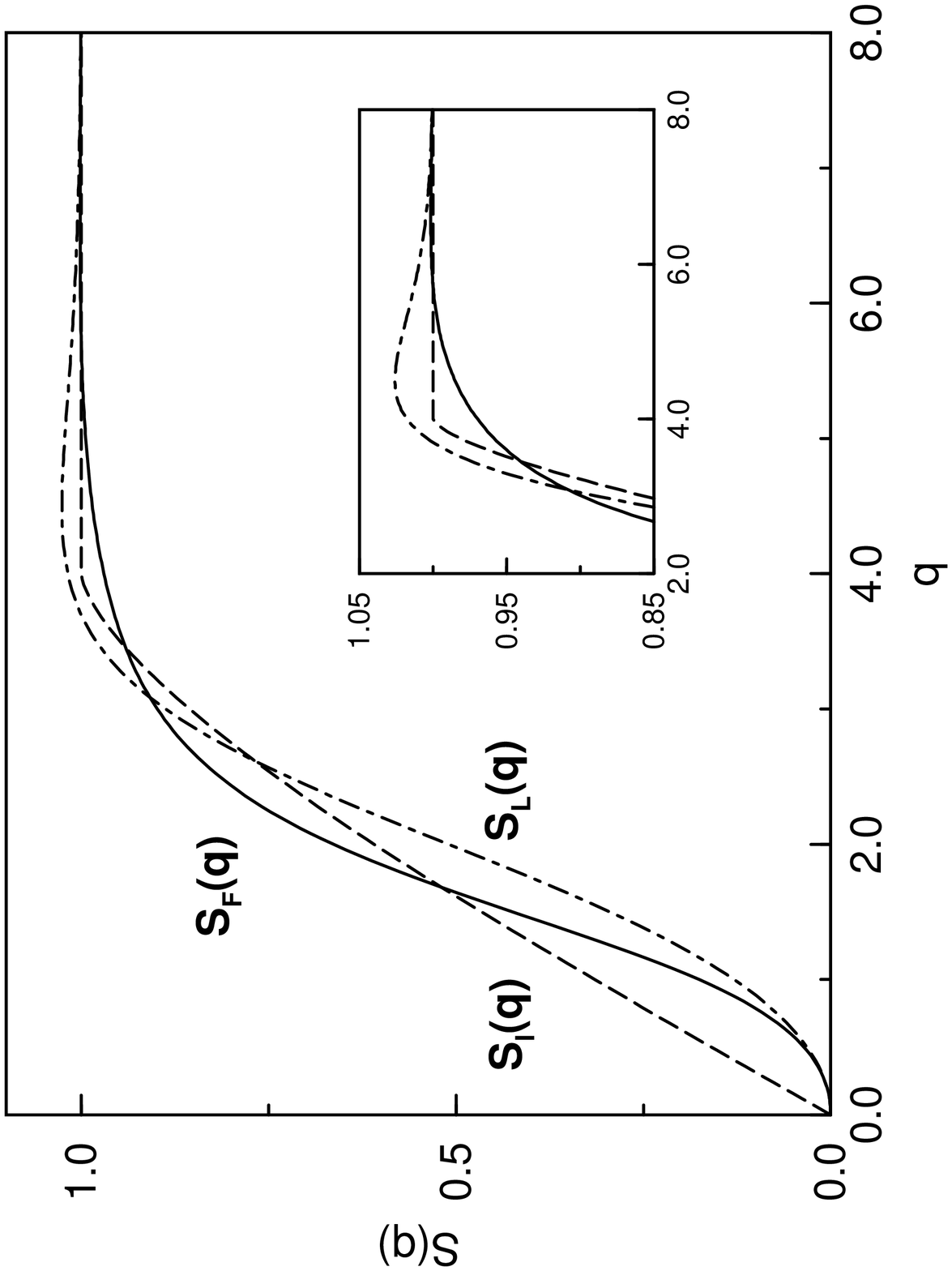}}
  \label{fig2}
\end{picture}
\end{center}
\vspace*{1.2cm}

\noindent
FIG. 2: The static structure factor $S(q)$ as a function of $q=kR$, for the 
non-interacting system $[S_I(q)]$ [Eq. (8)], the Laughlin state (1) for a Bose 
system $(S_L(q)]$, and the ``Fermi'' state (3) $S_F(q)$. A blowup of the region 
around unity is given as inset.
\vspace*{1.0cm}

Given these two functions, $u_I(r)$ now can be obtained from inverting the 
hypernetted-chain (HNC) equations \cite{springer,book},
\begin{equation}
u_I(r)=-\ln g_I(r)+\frac1{n_e}{\cal F}\left\{\frac{\left[S_I(k)-1\right]^2}
{S_I(k)}\right\},
\end{equation}
where $\cal F$ denotes the two-dimensional Fourier transform
\begin{eqnarray}
F(k) & = & 2\pi\int_0^\infty r dr F(r) J_0(kr)\nonumber \\
F(r) & = & \frac1{2\pi}\int_0^\infty k dk F(k) J_0(kr),\nonumber \\
\nonumber
\end{eqnarray}
where $J_0(kr)$ is the Bessel function of the first kind of order zero.

The implication of Eqs. (7)-(9) is that in the two-body level, $\phi(r)$ in 
Eq. (4) can be chosen such that we reproduce the exact two-body radial 
distribution function and the static structure factor corresponding to the 
{\it full determinant}. We note that $u_I(r)$ is a numerically strictly 
decreasing function, but it is a long-ranged function (because for small $q$, 
$\tilde{u}_I(q)\equiv{\cal F}[u_I(r)]\sim q^{-1}$). Therefore one needs to take 
special care about the long- and short-range behavior of 
$u_t(r)$ \cite{springer} while solving the HNC equations for the one-component 
plasma. In our numerical calculations we have used the dimensionless variables
$x=r/R$ and $q=kR$ where $R=\sqrt{2m} \ell_0$ is the ion-disk radius 
\cite{book}. The method of deriving the HNC equations for $u_t(x)$ is similar 
to that for $u(x)$ and with proper choice of the long- and short-range 
functions \cite{book,springer} a numerically rapidly convergent set of 
equations are obtained which lead to $g(x)$ and its Fourier transform, $S(q)$. 
In the case of 
$u_I(x)=0$, the pair-correlation function $g_L(x)$ for the Laughlin state is 
plotted in Fig. 1. The ground state energy coresponding to that state is 
$E_L=-0.480 e^2/\epsilon\ell_0$. The pair-correlation function
corresponding to state (3) is denoted by $g_F(x)$ in Fig. 1. Clearly, the
``Fermi hole'' is not much affected by the introduction of the Laughlin
correlation function, but marked deviations of $g_F(x)$ from $g_I(x)$ occur 
near the maximum of $g_L(x)$.

The static structure functions $S(q)$ vs $q$ for the various cases are shown 
in Fig. 2. As with the pair-correlation functions, $S_I(q)$ corresponds to
the ideal system result, $S_F(q)$ is the present result for the Fermi-fluid
state (3) at $\nu=\frac12$ and $S_L(q)$ is the structure function for the
Laughlin state (1). For small $q$, it has been predicted that $S(q)\propto 
q^2$ in the Fermi fluid at $\nu=\frac12$ \cite{duncan}. Similar behavior is 
observed here for $S_F(q)$. The difference between the present results and 
the Laughlin results, $g_{diff}=g_F(x)-g_L(x)$ and $S_{diff}=S_F(q)-S_L(q)$, 
are plotted in Fig. 3. These are oscillatory functions with a rapidly decreasing
amplitude. In finite-size system calculations \cite{rezayi}, a sinusoidal 
oscillation in $g_{diff}(x)$ was taken as an indication of a Fermi fluid 
behavior. We note that the correlation functions for the non-ideal system
show much less oscillations around unity in accordance with the properties
of a uniform density fluid (inset of Fig. 1) and therefore the difference in 
correlation functions is also rapidly damped. Interestingly, $S_{diff}(k)$
develops a positive peak slightly below $k\sim2k_F$ and a negative peak beyond
that $k$. Finally, we find the ground state energy for the state (3) and for 
the Coulomb potential to be $E_F=-0.448\,e^2/\epsilon\ell_0$, which is very 
different from the energy of the Laughlin state $E_L$ but very close to the 
energy value $E_0$, extrapolated for an infinite system from the finite-size 
system results mentioned in the introduction. This agreement between the energy 
of the state (3) and the estimate $E_0$ is a strong indication that our 
Fermi-liquid description has the right correlations and correct statistics 
needed to describe a Fermi liquid behavior at $\nu=\frac12$.

\begin{center}
\begin{picture}(200,200)
\put(0,0){\includegraphics{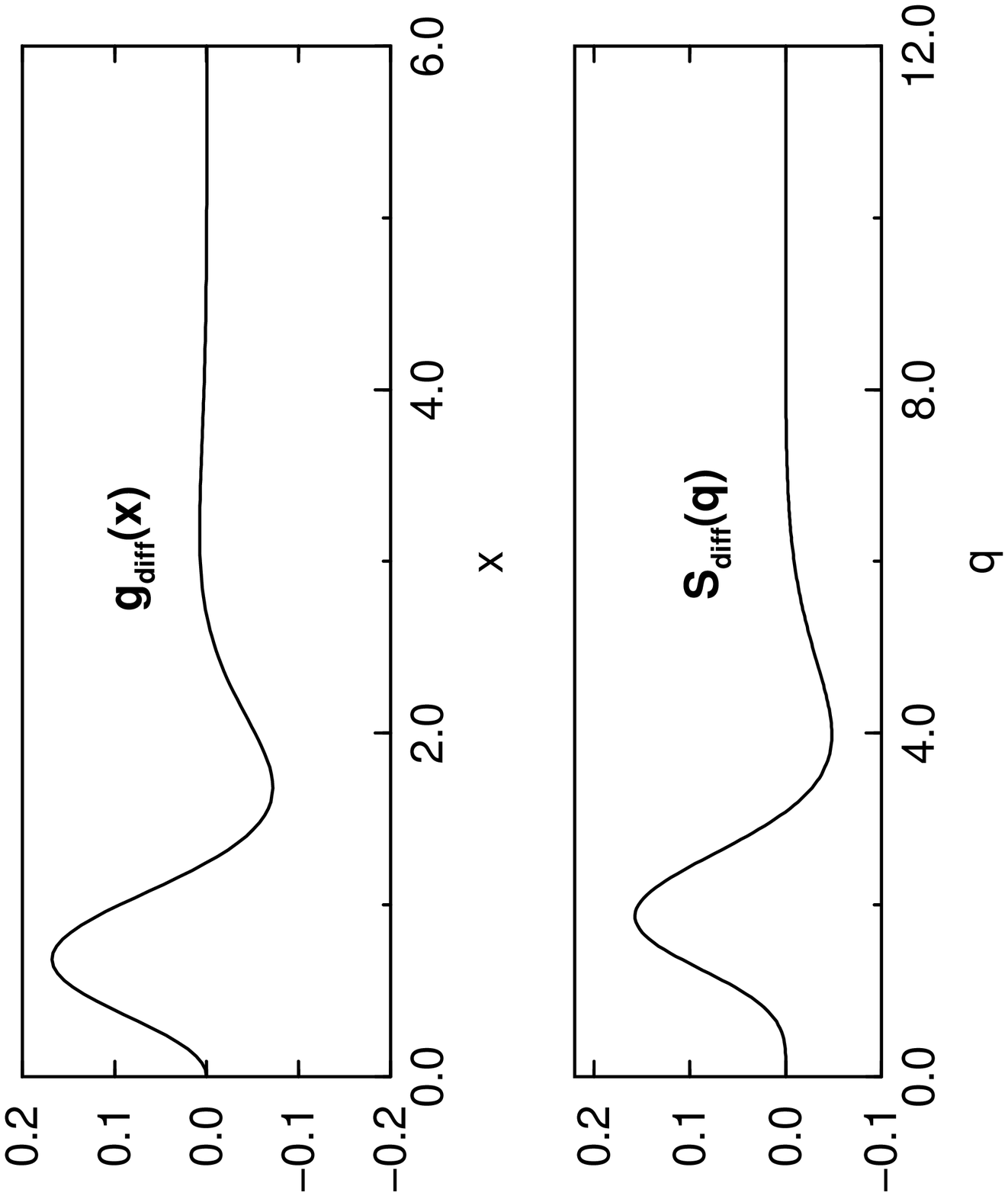}}
\label{fig3}
\end{picture}
\end{center}
\vspace*{.5cm}

\noindent
FIG. 3: The difference between the ``Fermi'' and ``Bose'' functions (a)
$g_{diff}(x)$ vs $x$ and (b) $S_{diff}(q)$ vs $q$.
\vspace*{1.0cm}

In summary, our simplified choice (3) of the ground state wave function for the
Fermi fluid state at $\nu=\frac12$ has led to a microscopic approach where
we can calculate the physical quantities in the thermodynamic limit. The
pair-correlation function, the structure function, and the ground state energy 
are in good agreement with the expected behavior at $\nu=\frac12$. This approach
can also be suitably modified to calculate the one-body density matrix and
the nature of the off-diagonal long-range order (ODLRO)
\cite{suther}, which will provide more information about the Fermi nature of 
the proposed state. In defense of our choice of (3), we should mention
that the wave function which would result from the operation of ${\cal
P}_{LLL}$ will be a wave function in the lowest Landau level, and therefore be 
similar to the Laughlin-like wave function (but with correct statistics). Hence 
our choice of (5), which is formally similar to the Laughlin approach should be 
a suitable approximation for the full wave function (2). This is supported by 
our numerical results presented here.

I would like to thank Pekka Pietil\"ainen for helpful discussions. I also
thank Peter Fulde for kind hospitality.

\end{document}